# Diabetic Retinopathy Detection Based on Convolutional Neural Networks with SMOTE and CLAHE Techniques Applied to Fundus Images

Sidhiq Mardianta
*Faculty of Computer Science*
*Universitas Dian Nuswantoro*
Semarang, Indonesia
dhiqde@gmail.com

Affandy, Affandy
*Faculty of Computer Science*
*Universitas Dian Nuswantoro*
Semarang, Indonesia
affandy@dsn.dinus.ac.id

Catur Supriyanto
*Faculty of Computer Science*,
*Indonesia Dinus Research Group for*
*AI in Medical Science* (DREAMS)
*Universitas Dian Nuswantoro*
Semarang, Indonesia
catur.supriyanto@dsn.dinus.ac.id

Moch Arief Soeleman
*Faculty of Computer Science*
*Universitas Dian Nuswantoro*
Semarang, Indonesia
arief22208@gmail.com

Adi Wijaya
*Department of Health Information*
*Management*
*Universitas Indonesia Maju*
Jakarta, Indonesia
adiwijaya@mail.ugm.ac.id

*Abstract*—Diabetic retinopathy (DR) is one of the major complications in diabetic patients' eyes, potentially leading to permanent blindness if not detected timely. This study aims to evaluate the accuracy of artificial intelligence (AI) in diagnosing DR. The method employed is the Synthetic Minority Over-sampling Technique (SMOTE) algorithm, applied to identify DR and its severity stages from fundus images using the public dataset "APTOS 2019 Blindness Detection." Literature was reviewed via ScienceDirect, ResearchGate, Google Scholar, and IEEE Xplore. Classification results using Convolutional Neural Network (CNN) showed the best performance for the binary classes normal (0) and DR (1) with an accuracy of 99.55%, precision of 99.54%, recall of 99.54%, and F1-score of 99.54%. For the multiclass classification No_DR (0), Mild (1), Moderate (2), Severe (3), Proliferate_DR (4), the accuracy was 95.26%, precision 95.26%, recall 95.17%, and F1-score 95.23%. Evaluation using the confusion matrix yielded results of 99.68% for binary classification and 96.65% for multiclass. This study highlights the significant potential in enhancing the accuracy of DR diagnosis compared to traditional human analysis

## I. Introduction

Diabetic Retinopathy (DR) is one of the major complications in the eyes of diabetic patients, potentially leading to permanent blindness if not detected in a timely manner. The International Diabetes Federation (IDF) reports that the number of people with diabetes is projected to reach 700 million by 2045 [1]. DR occurs when blood vessels in the retina swell and leak due to high blood sugar levels, causing vision impairment, heart attacks, kidney failure, and strokes [2]. Retinal diseases have become a leading cause of blindness in children worldwide. Identifying this disease is particularly challenging due to the variety of conditions affecting the retina fig.1 [3].

Early detection of (DR) is crucial, as it can prevent many patients from progressing to permanent blindness. Artificial intelligence (AI) has demonstrated higher accuracy in detecting DR compared to human analysis. In traditional deep learning models, cross-entropy is commonly employed as a loss function in one-stage end-to-end training methods. AI technology also offers the potential to reduce healthcare costs associated with severe diabetic retinopathies and to broaden the scope of screening. However, challenges such as class imbalance in data remain, which is particularly prevalent in medical diagnosis, including the diagnosis of diabetic patients. To address this, predicting performance scores in data mining is essential. basic concepts of data mining, emphasizing the importance of analyzing very large data sets, basic methodologies, and various techniques used in the field, including association rule mining, classification, regression, and clustering [4]. Research on classifying data in unbalanced classes has been extensively conducted, with some studies proposing the use of the k-Nearest Neighbor algorithm as an effective approach [5].

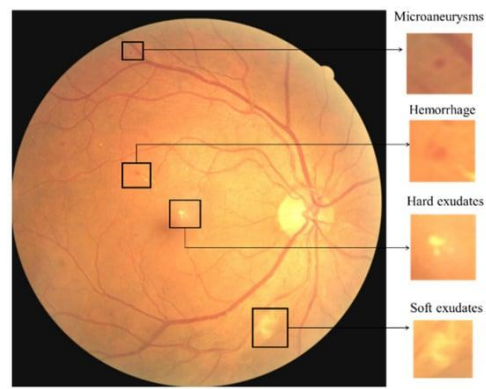

Fig. 1. Fundus image with variety conditions.

The use of convolutional neural networks (CNNs) in medical image classification is highlighted in this review. It covers issues with the diagnosis of DR due to class imbalance in the training data [6]. Additionally, using transfer learning techniques from pre-trained models requires adaptation to the characteristics of medical images. To address these issues, this research employs the Synthetic Minority Over-sampling Technique (SMOTE) and Contrast Limited Adaptive Histogram Equalization (CLAHE) to enhance fundus image quality and tackle data imbalance. Using the public dataset "APTOS 2019 Blindness Detection," this study aims to explore whether preprocessing techniques such as SMOTE and CLAHE can improve the accuracy of DR severity classification using a pre-trained CNN Xception model. Evaluation using t-SNE to visualize the imaging space indicates that this approach can assist medical professionals in diagnosing DR more effectively, as well as contribute to the understanding of artificial intelligence applications in the medical field.

## II. RELATED WORK

Recent studies in diabetic retinopathy (DR) classification have explored various deep learning-based approaches using the Kaggle APTOS 2019 dataset. Some of the methods that have been applied include the use of pre-trained CNNs for fundoscopic representation with spatial pooling techniques, which achieved 97.82% accuracy for binary classification [7]. Meanwhile, another study proposed a DenseNet121 model with certain modifications, resulting in 94.44% accuracy and 87% recall score [8]. A multimodal fusion model combining features from VGG16 and Xception has also been tested, achieving 96.10% accuracy for DR identification and 80.96% for severity classification [9]. A hybrid model based on VGG16 and Capsule Network used in another study resulted in 97.05% accuracy for DR identification and 75.50% for five-stage classification [10]. A Neural Architecture Search Network (NASNet)-based approach combined with t-SNE and v-SVM achieved 77.90% accuracy for multiclass classification [11]. In addition, a pre-trained Inception-ResNet-v2-based model achieved 82.18% accuracy for five-class classification [12]. Another study used transfer learning with multilayer perceptron (MLP) to classify five-stage DR from fundus images [13], while a customized CNN model with five convolutional layers achieved 77% accuracy for five-stage DR classification [14].

## III. PROPOSED METHOD

In this study, the dataset was processed using the Convolutional Neural Network (CNN) method. This study proposes the use of Synthetic Minority Oversampling Technique (SMOTE) during the preprocessing phase to address the problem of data imbalance. This approach was employed to enhance image quality prior to the training phase [8].

In previous studies, the CLAHE method was utilized to reduce noise by defining a kernel matrix. This approach involved replacing the intensity values of each pixel in the input image. Furthermore, researchers incorporated the SMOTE prior to classifying the dataset using a CNN [7].

The following are the methods proposed by researchers in this research Fig. 2.

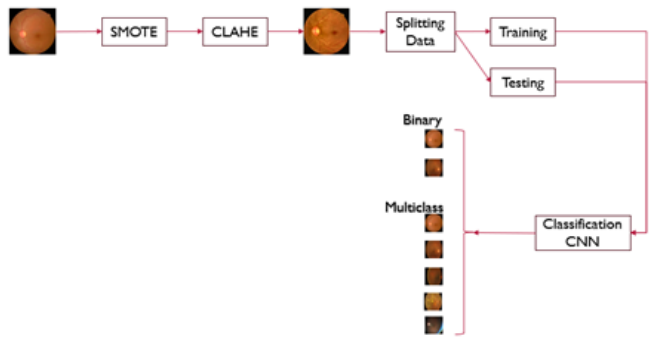

Fig. 2. The proposed framework for the identification of DR.

### A. Data Collection Method

The study on (DR) detection utilizes a publicly available dataset sourced from Kaggle [9]. This dataset, comprising over 88,000 publicly accessible images captured using various cameras at diverse angles and dimensions, is considered one of the most significant resources for DR research. The dataset is divided into 40% for training and 60% for testing purposes. Given the diverse range of cameras employed, the dataset exhibits varying levels of image quality. As highlighted by Wilkinson et al [10], the dataset adopts a five-class annotation system. However, the rare DR severity levels (classes 3 and 4) account for less than 5% of the total dataset, resulting in a significant class imbalance.

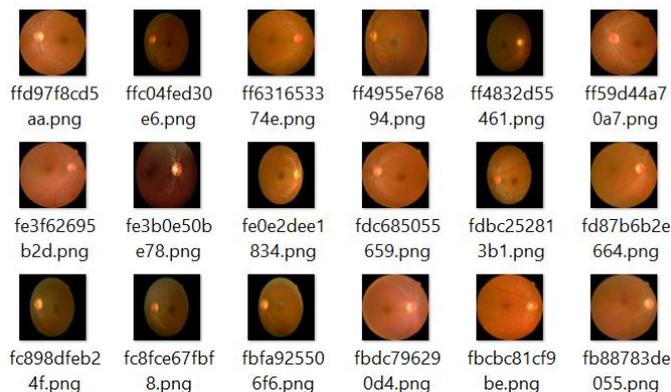

Fig. 3. Dataset aptos 2019 blindness detection.

### B. Data Preparations

In this study, DR is categorized into multiple classes using an image dataset, where the dataset is pre-organized into subfolders based on the desired categories: normal, mild, moderate, severe, and proliferative:

0 - No DR

1 - Mild

2 - Moderate

3 - Severe

4 - Proliferative DR

### C. Handling Class Problems

DR detection often encounters the challenge of class imbalance, where higher retinopathy severity levels have significantly fewer samples compared to the normal class. Researchers have considered employing techniques such as SMOTE to generate synthetic samples for the higher severity classes, thereby balancing the dataset.

### D. Data Preprocessing

The data preprocessing in this study involves several stages, including the following steps:

1. Data Augmentation: Enhancing the training dataset by applying transformations such as rescaling (normalization), zooming, horizontal and vertical shifts, horizontal flipping, rotation, and shearing. These techniques increase the diversity of the training data, improving model robustness.

2. Data Splitting: The dataset is separated into subsets for training and validation, usually with a training to validation ratio of 80:20 and a validation ratio of 20:40.

This makes it easier to monitor and assess the model while it is being trained.

3. Synthetic Oversampling using SMOTE: SMOTE is applied to oversample minority classes in the training dataset, addressing class imbalance.

To handle data imbalance effectively, SMOTE is frequently used to increase the number of samples in minority classes [11]. The following steps outline the SMOTE implementation:

1. Minority Class Identification: Identify minority classes in the dataset, i.e., those with fewer samples.

2. Selection of Minority Samples: Select one or more samples from the minority class as initial points for generating new synthetic samples.

3. Distance Calculation: Determine the separation between the minority samples that were chosen and the other samples in the minority class.

4. Neighbor Selection: Choose several nearest neighbors of the selected minority samples based on the calculated distances.

5. Synthetic Sample Generation: For each selected neighbor, determine a proportion (usually between 0 and 1) to create new synthetic samples along the line connecting the selected minority sample to its neighbor. For instance, if the proportion is 0.5, the midpoint of the line between two samples will be the location for the new synthetic sample.

6. Process Iteration: Repeat steps 2 through 5 for multiple minority samples until the desired number of synthetic samples is generated.

7. Integration with the Original Dataset: Combine the newly generated synthetic samples with the original dataset.

8. Model Training: Train the machine learning model on the updated dataset.

*E. CLAHE (Contrast Limited Adaptive Histogram Equalization)*

The significance of image preprocessing prior to analysis is paramount in achieving more accurate predictions using machine learning. Various techniques have been developed to enhance the quality of medical images utilized in disease detection through machine learning approaches. One commonly employed technique is CLAHE, specifically designed to improve the quality of medical images [12]. Low-contrast medical images hold potential for broader applications. CLAHE serves as an alternative implementation of Adaptive Histogram Equalization (AHE).

Unlike traditional Histogram Equalization (HE), which treats the entire image as a single entity for equalization, CLAHE divides the image into smaller regions known as tiles, applying AHE individually to each tile. This method limits the amplification of contrast in CLAHE by clipping the histogram at a user-defined threshold, referred to as the clip limit. The clipping level determines the extent to which noise in the histogram is reduced, thereby enhancing the contrast achieved through CLAHE. Each image undergoes localized contrast enhancement by adapting the local histogram across different regions of fundus images [13].

*F. CNN Training*

In this study, the dataset was divided into training and validation sets. CNN was designed and trained to identify retinopathy patterns from retinal images. The performance of the model is largely dependent on the choice of a suitable CNN architecture, as well as the efficient use of convolutional layers, max pooling, and fully connected layers.

*G. Validation and Evaluation*

The model is evaluated on the validation/test set to assess its performance and ability to classify new images. Subsequently, To evaluate the model's performance, evaluation measures including accuracy, precision, recall, and F1-score are computed.

*H. Confusion Matrix*

A technique for assessing classification models' performance, the confusion matrix gauges how well the model predicts the target class. It is commonly used in classification problems where data is assigned to specific classes, such as positive and negative classes. The confusion matrix serves as the foundation for calculating several important evaluation metrics in classification analysis, including accuracy, precision, recall (sensitivity), F1-score, and others [15]. As shown in Figure 5, the confusion matrix contains four key elements:

1. True Positives (TP): The samples that actually belong to the positive class (target) and are accurately predicted as positive by the model. In medical words, they may be individuals that are genuinely afflicted with the disease and are appropriately diagnosed as such by the model.

2. True Negatives (TN): These are the samples that genuinely belong to the negative class (non-target) and are accurately predicted by the model to belong to the negative class. In medical terms, these are healthy patients who are correctly identified as healthy by the model.

3. False Positives (FP): These are the samples that are incorrectly predicted by the model to belong to the positive class, even though they actually belong to the negative class (non-target). This is also known as a Type I error.

4. False Negatives (FN): These are the samples that truly belong to the positive class (target), but are incorrectly predicted by the model to belong to the negative class. This is also known as a Type II error.

|  | Actual Values | |
|---|---|---|
|  | Positive (1) | Negative (0) |
| Predicted Values — Positive (1) | TP | FP |
| Predicted Values — Negative (0) | FN | TN |

Fig. 4. Confusion matrix.

Using these elements, various evaluation metrics can be calculated as follows:

1. Accuracy

The proportion of correct predictions, calculated as TP + TN) / (TP + TN + FP + FN). This metric evaluates the overall ability of the model to classify the data accurately.

$$Accuracy = \frac{TP+TN}{TP+TN+FP+FN} \quad (1)$$

2. Precision

The ratio of true positive predictions to the total predicted positives, calculated as TP/(TP+FP). Precision measures the model's ability to avoid false positive predictions.

$$Precision = \frac{TP}{TP+FP} \quad (2)$$

3. Recall or Sensitivity

The proportion of actual positive cases correctly predicted by the model, calculated as TP / (TP + FN). Recall assesses the model's ability to identify all true positive cases.

$$Sensitivity = \frac{TP}{TP+FN} \quad (3)$$

4. Specificity (True Negative Rate)

The proportion of actual negative cases correctly predicted by the model, calculated as TN / (TN + FP). Specificity evaluates the model's ability to identify all true negative cases.

$$Specificy = \frac{TN}{TN+FP} \quad (4)$$

5. F1-Score

A metric that combines precision and recall to provide a comprehensive view of the model's performance, calculated as 2 * (Precision * Recall) / (Precision + Recall).

$$F1 - Score = \frac{2*(Precision*Recall)}{Precision+Recall} \quad (5)$$

AUC-ROC is a numerical metric that quantifies the model's ability to distinguish between two classes. The AUC-ROC score ranges from 0 to 1, with a value of 1 denoting exceptional class separation and a result of 0.5 denoting no better performance than random guessing. Higher AUC-ROC values signify better model performance in correctly predicting positive classes while avoiding false predictions for negative classes.

## IV. EXPERIMENT RESULT

In this study, the dataset used consists of retinal fundus images obtained from the journal Computers in Biology and Medicine [7], which is publicly available on the Kaggle platform. The collected data is categorized based on the severity of diabetic retinopathy into several classes. The binary classification consists of 1,805 images labeled as Normal and 1,857 images labeled as DR. For the multi-class classification, the data is divided into five subsets: NO_DR, Mild (370 images), Moderate (999 images), Severe (193 images), and Proliferative_DR (295 images).

The dataset collection serves as a basis for research in analyzing and evaluating models for diabetic retinopathy severity classification. With high-quality data, the study aims to contribute significantly to early detection of the disease. As shown Fig. 5, illustrates the dataset prior to CLAHE preprocessing.

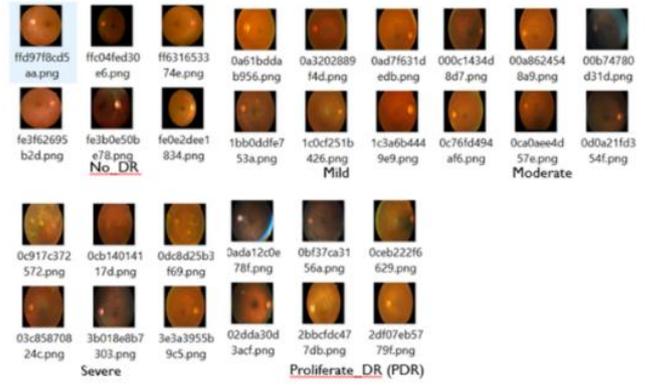

Fig. 5. Fundus image dataset before preprocessing.

### A. Preprocessing Data

The data preprocessing process involved the application of SMOTE to address class imbalance and CLAHE to enhance image quality. SMOTE successfully generated oversampled data for minority classes, while CLAHE was applied to the luminance component of the images to improve contrast and detail fig. 6, these preprocessing steps are critical to ensuring that the model can effectively learn from high-quality and balanced data.

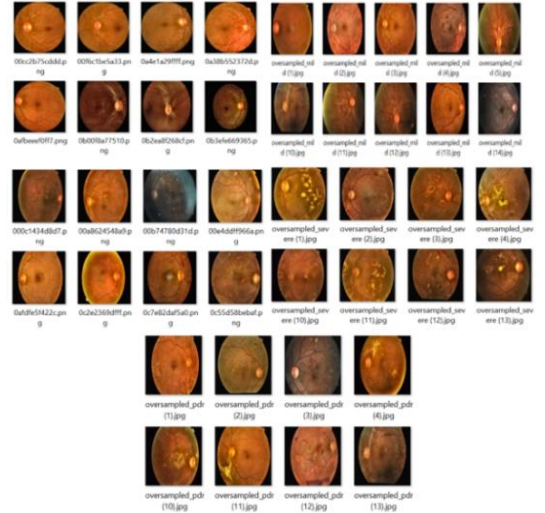

Fig. 6. Fundus image preprocessing CLAHE results.

### B. Preprocessing CNN Classification by Dividing Data

In binary classification, 80% of the dataset was set aside for training and 20% for testing, whereas in multiclass classification, 85% was set aside for training and 15% for testing. The train_test_split function from the scikit-learn package was used to carry out this split, which ensures that the class proportions from the original dataset are preserved in the resulting subsets. While the test sets comprise 2,198 and 807 pictures, respectively, the final distribution contains 8,788 training samples for binary classification and 4,571 for multiclass classification.

### C. Classification and Evaluation Models

The classification model employed in this study is a CNN comprising multiple convolutional layers, max pooling layers, and dense layers, with L2 regularization and dropout to prevent overfitting. The model was trained using various optimizers and evaluated based on accuracy, precision, and recall metrics. The evaluation's findings show that the model's accuracy for binary and multiclass classification was 99.55% and 95.26%,

respectively. The table below summarizes the model's evaluation metrics:

TABEL I. SUMMARIZES MODEL EVALUATION METRICS

| Evaluation Metrics | Binary class | Multi class |
|---|---|---|
| Accuracy (%) | 99.55 | 95.26 |
| Precision (%) | 99.55 | 95.26 |
| Recall (%) | 99.55 | 95.17 |
| FI-score (%) | 99.54 | 95.23 |
| AUC (%) | 99.98 | 99.64 |

The table above demonstrates that the proposed model not only achieves high accuracy but also exhibits consistent performance in terms of precision and recall, which are critical in the context of medical diagnosis.

The findings of this study reveal that the combination of SMOTE, CLAHE, and CNN significantly enhances model performance in detecting diabetic retinopathy. With exceptionally high accuracy 99.55% accuracy, precision, and recall for binary classification, and 95.26% accuracy, 95.26% precision, and 95.17% recall for multiclass classification this research makes a substantial contribution to the development of early detection methods for diabetic retinopathy. These results pave the way for future research to explore factors influencing model performance and develop more advanced techniques to further improve the accuracy and efficiency of disease diagnosis.

TABEL II. COMPARISON PERFORMANCE WITH OTHER METHODS

| Methods | Accuracy | |
|---|---|---|
| | Binary | Multiclass |
| DNN,CNN [7] | 97.82% | 80.96% |
| DenseNet121 [8] | 94.44% | - |
| Xception,Inception, MobileNet,ResNet50 + DNN [9] | 96.10% | - |
| VGG16+DRISTI [10] | 97.05% | 75.50% |
| CNN, NASNet, t-SNE [11] | - | 77.90% |
| CNN,ResNet-v2 [12] | - | 82.18% |
| CNN, Transfer Learning [14] | - | 77.00% |
| This Study | 99.55% | 99.55% |

Table 2 compares the performance of several deep learning models for the categorization of diabetic retinopathy (DR). The suggested model performs better than earlier research that used CNN, DenseNet121, and hybrid models, despite the fact that they all achieved excellent accuracy. The suggested model outperformed previous approaches and showed improved dependability in DR detection by combining SMOTE and CLAHE, achieving 99.55% accuracy for both binary and multiclass classification.

V. CONCLUSION

This study demonstrates that combining image preprocessing techniques such as SMOTE (Synthetic Minority Over-sampling Technique) and CLAHE (Contrast Limited Adaptive Histogram Equalization) with Convolutional Neural Network (CNN) classification significantly improves the accuracy of diabetic retinopathy detection. The use of SMOTE addresses class imbalance, while CLAHE enhances image quality, leading to better model performance. The CNN model achieved high accuracy in both binary and multi-class classification, with evaluation metrics like the Confusion Matrix and ROC AUC Score confirming its ability to distinguish between classes effectively.

The results highlight that the proposed model offers strong potential for early detection and monitoring of diabetic retinopathy. These findings suggest that integrating advanced preprocessing techniques with CNN classification could be an effective approach for developing reliable and accurate diagnostic systems, providing valuable support for healthcare professionals in delivering timely and accurate care to patients.

REFERENCES

[1] P. S. Suvi Karuranga, Belma Malanda, Pouya Saeedi, "IDF Diabetes Atlas 9th edition," 2019, [Online]. Available: https://www.diabetesatlas.org

[2] K. Boyd, "American Academy of Ophthalmology-What is Diabetic Retinopathy," *Accessed Sep*, vol. 10, p. 2021, 2020.

[3] W. L. Alyoubi, M. F. Abulkhair, and W. M. Shalash, "Diabetic retinopathy fundus image classification and lesions localization system using deep learning," *Sensors*, vol. 21, no. 11, Jun. 2021, doi: 10.3390/s21113704.

[4] D. P, P. Nayak, P. Poojary, P. Talekar, and Prassana, "Data Mining: Concepts, Techniques, and Applications," *Int. J. Adv. Res. Sci. Commun. Technol.*, pp. 120–128, Nov. 2024, doi: 10.48175/IJARSCT-22821.

[5] S. Mutrofin, A. Mu'alif, R. V. H. Ginardi, and C. Fatichah, "Solution of class imbalance of k-nearest neighbor for data of new student admission selection," *Int. J. Artif. Intell. Res.*, 2019, [Online]. Available: https://api.semanticscholar.org/CorpusID:199131273

[6] F. A. Mohammed, K. K. Tune, B. G. Assefa, M. Jett, and S. Muhie, "Medical Image Classifications Using Convolutional Neural Networks: A Survey of Current Methods and Statistical Modeling of the Literature," *Mach. Learn. Knowl. Extr.*, vol. 6, no. 1, pp. 699–736, Mar. 2024, doi: 10.3390/make6010033.

[7] J. D. Bodapati, N. S. Shaik, and V. Naralasetti, "Diabetic Retinopathy (DR) is a micro vascular complication caused by long-term diabetes mellitus. Unidentified diabetic retinopathy leads to permanent blindness. Early identification of this disease requires frequent complex diagnostic proce-dure which is e," *J. Ambient Intell. Humaniz. Comput.*, vol. 12, no. 10, pp. 9825–9839, Oct. 2021, doi: 10.1007/s12652-020-02727-z.

[8] S. Chaturvedi, K. Gupta, V. Ninawe, and P. Prasad, *Automated Diabetic Retinopathy Grading using Deep Convolutional Neural Network*. 2020.

[9] J. D. Bodapati *et al.*, "Blended multi-modal deep convnet features for diabetic retinopathy severity prediction," *Electron.*, vol. 9, no. 6, Jun. 2020, doi: 10.3390/electronics9060914.

[10] G. Kumar, S. Chatterjee, and C. Chattopadhyay, "DRISTI: a hybrid deep neural network for diabetic retinopathy diagnosis," *Signal, Image Video Process.*, vol. 15, Nov. 2021, doi: 10.1007/s11760-021-01904-7.

[11] V. Dondeti, J. D. Bodapati, S. N. Shareef, and V. Naralasetti, "Deep convolution features in non-linear embedding space for fundus image classification," *Rev. d'Intelligence Artif.*, vol. 34, no. 3, pp. 307–313, Jun. 2020, doi: 10.18280/ria.340308.

[12] A. K. Gangwar and V. Ravi, "Diabetic Retinopathy Detection Using Transfer Learning and Deep Learning," *Adv. Intell. Syst. Comput.*, vol. 1176, pp. 679–689, 2021, doi: 10.1007/978-981-15-5788-0_64.

[13] S. H. Kassani, P. Hosseinzadeh Kassani, R. Khazaeinezhad, M. Wesolowski, K. Schneider, and R. Deters, *Diabetic Retinopathy Classification Using a Modified Xception Architecture*. 2019. doi: 10.1109/ISSPIT47144.2019.9001846.

[14] O. Dekhil, A. Naglah, M. Shaban, M. Ghazal, F. Taher, and A. Elbaz, "Deep Learning Based Method for Computer Aided Diagnosis of Diabetic Retinopathy," *IST 2019 - IEEE Int. Conf. Imaging Syst. Tech. Proc.*, Dec. 2019, doi: 10.1109/IST48021.2019.9010333.

[15] J. Görtler *et al.*, *Neo: Generalizing Confusion Matrix Visualization to Hierarchical and Multi-Output Labels*. 2022. doi: 10.1145/3491102.3501823.